\documentclass[fleqn,usenatbib]{mnras}

\usepackage{newtxtext,newtxmath}

\usepackage[T1]{fontenc}
\usepackage{ae,aecompl}

\usepackage{graphicx}	
\usepackage{amsmath}	
\usepackage{amssymb}	
\usepackage[FIGTOPCAP]{subfigure}

\def\f{\frac}
\def\nn{\nonumber}
\def\ln{\mathrm{ln}}

\def\cov{\mathbf{C}}

\def\data{\mathbf{d}}

\def\btheta{\boldsymbol{\theta}}
\def\bphi{\boldsymbol{\phi}}

\def\t{\mathbf{t}}
\def\L{\mathcal{L}}

\title[Nuisance hardened data compression for fast likelihood-free inference]{Nuisance hardened data compression for fast likelihood-free inference}

\author[J. Alsing, B. Wandelt]{
Justin Alsing$^{1,2,3}$\thanks{E-mail: justin.alsing@fysik.su.se}
and Benjamin Wandelt$^{2,4}$
\\
$^{1}$Oskar Klein Centre for Cosmoparticle Physics, Stockholm University, Stockholm SE-106 91, Sweden\\
$^{2}$Center for Computational Astrophysics, Flatiron Institute, 162 5th Ave, New York City, NY 10010, USA\\
$^{3}$Imperial Centre for Inference and Cosmology, Department of Physics, Imperial College London, Blackett Laboratory,\\  Prince Consort
Road, London SW7 2AZ, UK\\ 
$^{4}$Sorbonne Universit\'e, Institut Lagrange de Paris (ILP), 98 bis boulevard Arago, F-75014 Paris, France\\
}

\date{Accepted XXX. Received YYY; in original form ZZZ}

\pubyear{2018}

\begin{document}
\label{firstpage}
\pagerange{\pageref{firstpage}--\pageref{lastpage}}
\maketitle

\begin{abstract}
In this paper we show how nuisance parameter marginalized posteriors can be inferred directly from simulations in a likelihood-free setting, without having to jointly infer the higher-dimensional interesting and nuisance parameter posterior first and marginalize a posteriori. The result is that for an inference task with a given number of interesting parameters, the number of simulations required to perform likelihood-free inference can be kept (roughly) the same irrespective of the number of additional nuisances to be marginalized over. To achieve this we introduce two extensions to the standard likelihood-free inference set-up. Firstly we show how nuisance parameters can be re-cast as latent variables and hence automatically marginalized over in the likelihood-free framework. Secondly, we derive an asymptotically optimal compression from $N$ data down to $n$ summaries -- one per interesting parameter -- such that the Fisher information is (asymptotically) preserved, but the summaries are insensitive (to leading order) to the nuisance parameters. This means that the nuisance marginalized inference task involves learning $n$ interesting parameters from $n$ ``nuisance hardened" data summaries, regardless of the presence or number of additional nuisance parameters to be marginalized over. We validate our approach on two examples from cosmology: supernovae and weak lensing data analyses with nuisance parameterized systematics. For the supernova problem, high-fidelity posterior inference of $\Omega_m$ and $w_0$ (marginalized over systematics) can be obtained from just a few hundred data simulations. For the weak lensing problem, six cosmological parameters can be inferred from just $\mathcal{O}(10^3)$ simulations, irrespective of whether ten additional nuisance parameters are included in the problem or not. If needed, an approximate posterior for the nuisance parameters can be re-constructed a posteriori as a pseudo-Blackwell-Rao estimator (without running any additional simulations).
\end{abstract}

\begin{keywords}
data analysis: methods
\end{keywords}


\section{Introduction}
\label{sec:introduction}
Likelihood-free inference is emerging as a new paradigm for performing principled inference from cosmological data using forward simulations only, incorporating exactly all known effects that can be successfully simulated without relying on approximate likelihoods \citep{Schafer2012,Cameron2012,Weyant2013,Robin2014,Lin2015,Ishida2015, Akeret2015,Jennings2016,Hahn2017,Kacprzak2017,Carassou2017,Davies2017,Alsing2018delfi,Alsing2019neural}.

In \citet{Alsing2018delfi,Alsing2019neural} we showed that likelihood-free inference is achievable from just $\mathcal{O}(10^3)$ forward simulations for typical problems in cosmology, with $n \sim 6$ or so cosmological parameters of interest. However, while many problems have a small number of interesting parameters, they often come with a large number of additional nuisance parameters that ultimately need marginalizing over. In the standard Bayesian approach, interesting and nuisance parameters are first inferred together and the resulting joint posterior is then marginalized over the nuisances a posteriori. For likelihood-free methods, solving the initially higher-dimensional inference task (over interesting and nuisance parameters together) would typically demand a much larger number of forward simulations, even though the nuisance parameters are later marginalized out anyway. When forward simulations are expensive, this may pose a critical impasse for likelihood-free analyses.

In this paper we show that in the likelihood-free setting, nuisance marginalized posteriors can be inferred directly from forward simulations, bypassing the higher-dimensional ``interesting plus nuisance parameters" inference task entirely. The result is that for problems with a given number of interesting parameters, the number of simulations required for performing likelihood-free inference is (roughly) the same irrespective of the number of additional nuisance parameters that need marginalizing over. For problems in cosmology and astrophysics where often a modest number of interesting parameters are accompanied by a large number of nuisances, this is a major step toward enabling likelihood-free analyses where simulations are expensive.

To enable direct inference of nuisance marginalized posteriors we introduce two extensions to the standard likelihood-free inference set-up. Firstly, we re-cast nuisance parameters as local latent variables in the forward simulations so they are automatically marginalized over with no additional approximations or loss of information. This way, unknown nuisances are effectively treated as an additional source of noise in the problem, allowing nuisance marginalized posteriors to be inferred directly.

Secondly, we derive an asymptotically optimal compression of $N$ data down to $n$ summaries -- one per interesting parameter -- such that the Fisher information is preserved, but the summaries are insensitive (to leading order) to the nuisance parameters. This allows us to keep the number of informative ``nuisance hardened" summary statistics at one per interesting parameter, irrespective of the number of nuisance parameters in the problem.

With nuisance hardened compressed summaries and nuisances cast as latent parameters, the nuisance marginalized likelihood-free inference task involves learning $n$ interesting parameters from data compressed down to $n$ informative summaries, regardless of the presence or number of additional nuisance parameters. The complexity and hence number of simulations required will then be (roughly) independent of the number of nuisances, expanding the range of problems that can be feasibly tackled with likelihood-free methods.

The plan of the paper is as follows. In \S \ref{sec:lfi} and \ref{sec:compression} we review the key ideas of likelihood-free inference and data-compression. In \S \ref{sec:nuisance_latent} we show how to re-cast nuisances as latent variables so they are automatically marginalized over in the likelihood-free setting. In \S \ref{sec:hardened} we derive an asymptotically optimal scheme for compressing data down to summary statistics that are hardened to nuisance parameters. In \S \ref{sec:jla}--\ref{sec:cosmic_shear} we validate the likelihood-free inference of nuisance marginalized posteriors on two case studies from cosmology: cosmological parameter inference from supernovae and cosmic shear data, with nuisance parameterized systematics. We conclude in \S \ref{sec:conclusions}.
\section{likelihood-free inference}
\label{sec:lfi}
The ideas presented in this paper are very general and can be applied to any likelihood-free inference approach. However, for simplicity we will frame the discussion in terms of density-estimation likelihood-free inference (DELFI; \citealp{Papamakarios2018,Lueckmann2018,Alsing2018delfi,Alsing2019neural}), which we briefly review below.

Suppose we have $p$ parameters $\boldsymbol{\phi}$ that we want to learn from data $\data$, that we have compressed down to some small number of informative summaries $\t\equiv \t(\data)$. In \S \ref{sec:compression} we will review two key methods for compressing large datasets down to $p$ summaries (that we will later extend to nuisance hardened data compression schemes), so throughout the paper we will also assume that $\t\in\mathbb{R}^p$.

Density-estimation likelihood-free inference proceeds in three steps:
\begin{enumerate}
\item Generate parameter-data summary pairs $\{\bphi, \t\}$ by running forward simulations,
\begin{align}
&\bphi \leftarrow q(\bphi),\nonumber \\
&\data \leftarrow \mathrm{simulator}(\data | \bphi), \nonumber \\
&\t = \mathrm{compression}(\data),
\end{align}
where $q(\bphi)$ is some parameter proposal distribution for choosing where to run simulations\footnote{The parameter proposal may be adaptively modified to optimize acquisition of relevant simulated parameter-data pairs \citep{Papamakarios2018}, or Bayesian optimization-style acquisition rules may be adopted for even more optimal simulation acquisition \citep{Lueckmann2018}.} and left-arrows indicate drawing random realizations.
\item Fit to these simulations $\{\bphi, \t\}$ a flexible parameterized model for the conditional density, $p(\t | \bphi)$ -- the sampling distribution of the summary statistics as a function of the parameters\footnote{There are three choices for fitting a density to $\{\bphi, \t\}$: a conditional density estimator for the posterior $p(\bphi | \t)$ \citep{Papamakarios2016,Lueckmann2017}, a model for the joint density $p(\bphi , \t)$ \citep{Alsing2018delfi}, or a conditional density estimator for the likelihood $p(\bphi | \t)$ \citep{Papamakarios2018,Lueckmann2018}. In this paper we will focus on the latter; see \citet{Alsing2019neural} and \citet{Papamakarios2018} for the relative merits of these three approaches.}. Mixture density networks (MDN) and masked autoregressive flows (MAF) have proven useful conditional density estimators in this context \citep{Papamakarios2016,Papamakarios2018,Alsing2019neural}.
\item Use the learned sampling density $p(\t | \bphi)$ evaluated at the observed data $\t_o$ in Bayes' theorem to obtain the (target) posterior:
\begin{align}
p(\bphi | \t_o) \propto p(\t_o | \bphi)p(\bphi).
\end{align}
\end{enumerate}
The full inference task is hence reduced to learning a $p$-dimensional density $p(\t | \bphi)$, as a function of $p$ parameters $\bphi$, from a set of simulated parameter-data pairs $\{\bphi, \t\}$. 

In order to enable fast likelihood-free inference, requiring the fewest simulations possible, we want to keep the dimensionality on the left and right and sides of the condition in $p(\t | \bphi)$ as low as possible for the problem we are trying to solve.
\section{Data compression and summary statistic choice}
\label{sec:compression}
When performing likelihood-free inference, it is usually necessary to compress large datasets down to a manageable number of informative summary statistics. This data compression should ideally be done so that is preserves as much of the information content of the full data (with respect to the parameters of interest) as possible. In this section we review three key data-compression schemes that summarize $N$ down to $p$ numbers -- one per parameter -- with the aim of preserving the (Fisher) information content of the data.
\subsection{Approximate score compression}
When the likelihood is known, the score function -- the gradient of the log-likelihood with respect to the parameters -- yields asymptotically optimal compression of $N$ data down to $p$ summaries \citep{Alsing2018}. Hence,
\begin{align}
\t = \nabla \L_*
\end{align}
represent a natural choice of summaries for parameters $\bphi$, where $\L$ denotes the log-likelihood, $*$ denotes evaluation about some fiducial parameters $\boldsymbol\phi_*$ and $\nabla\equiv\f{\partial}{\partial\bphi}$. When the gradient is taken close to the maximum-likelihood parameters, the summaries $\t$ saturate the information inequality, preserving the Fisher information content of the data. If a good expansion point is not available a priori, $\boldsymbol\phi_*$ can be iterated toward to the maximum likelihood for improved optimality if needed \citep{Alsing2018}.

Whilst the score presents a natural summary statistic choice for likelihood-free inference, it is only available if the likelihood-function is known. For likelihood-free applications where the likelihood is not known a priori, the score may still provide a guiding principle for choosing informative summaries. For many problems, an approximate likelihood (eg., Gaussian) may be used for the purposes of data compression, the only price being some loss of optimality. If no obvious likelihood approximation presents itself, once can learn the conditional density $p(\data | \btheta)$ from simulations locally in a small neighborhood around $\bphi_*$, which can then be used to define an approximate score. Alternatively, the score-function may be regressed directly from simulations \citet{Brehmer2018mining, Brehmer2018constraining, Brehmer2018guide}.

Note that for Gaussian data where the sensitivity to the model parameters is only in the mean, score-compression is equivalent to \textsc{moped} \citep{Heavens2000a}, and where the sensitivity to the parameters is only in the covariance it is equivalent to the optimal quadratic estimator \citep{Tegmark1997}.
\subsection{Neural network parameter estimators}
An emerging trend in cosmology is to find efficient parameter estimators from complex datasets by training deep neural networks to regress cosmological parameters from simulated data \citep{Ravanbakhsh2016,Schmelzle2017,Gupta2018,Ribli2018,Fluri2018}. The resulting trained neural networks can be viewed as radical data compression schemes, compressing large data sets down to a set of parameter estimators, whose sampling distributions are unknown. These neural network parameter estimators can be straightforwardly used in a subsequent likelihood-free analysis to obtain Bayesian posteriors from neural network summarized data sets.
\subsection{Information maximizing neural networks}
Combining the ideas of Fisher information preserving score-compression and neural network parameter estimators, Information Maximizing Neural Networks (\textsc{imnn}; \citealp{Charnock2018}) parameterize the data compression function $\t(\data):\mathbb{R}^N\rightarrow\mathbb{R}^p$ as a neural network, and train the network on a set of forward simulations of the data such that it maximizes the retained Fisher information content of the compressed summaries. The result is a likelihood-free Fisher optimal data compression scheme that is trained only on forward simulations, without any further (optimality-reducing) assumptions. In contrast to neural network parameter estimators, \textsc{imnn}s only need to run simulations around some fiducial model. This simplifies the learning task and has empirically been shown in test cases to give optimal compression with simple network architectures \citep{Charnock2018}.

In maximizing the Fisher information content of the summaries, \textsc{imnn}s also yield an estimate of the Fisher information matrix as a byproduct; this Fisher matrix estimate is essential for projecting nuisance parameters out of nuisance hardened summaries as we will see in \S \ref{sec:hardened}. 
\section{Likelihood-free inference of nuisance marginalized posteriors}
\label{sec:nuisance_latent}
Suppose now that our $p$ parameters $\boldsymbol{\phi} = (\boldsymbol{\theta}, \boldsymbol{\eta})$ constitute $n$ interesting parameters, $\boldsymbol{\theta}$, and $m$ nuisance parameters, $\boldsymbol{\eta}$.

When performing likelihood-free inference, parameter-data pairs $\{\bphi, \t\}$ are generated by running forward simulations:
\begin{align}
&\bphi \leftarrow q(\bphi),\nonumber \\
&\data \leftarrow \mathrm{simulator}(\data | \bphi), \nonumber \\
&\t = \mathrm{compression}(\data),
\end{align}
where $q(\bphi)$ is some proposal distribution. The ``simulator" is just a set of probabilistic statements that generate a realization of data $\data$ from some input (hyper-) parameters $\bphi$, via some intermediate local latent variable layers $\{\mathbf{z}\}$:
\begin{align}
&\mathrm{input\;parameters}\;\bphi,\nonumber \\
&\mathbf{z}_1 \leftarrow p(\mathbf{z}_1 | \bphi), \nonumber \\
&\mathbf{z}_2 \leftarrow p(\mathbf{z}_2 | \mathbf{z}_1, \bphi), \nonumber \\
&\vdots \nonumber \\
&\mathbf{z}_n \leftarrow p(\mathbf{z}_n | \mathbf{z}_1,\dots, \mathbf{z}_{n-1}, \bphi), \nonumber \\
&\data \leftarrow p(\data | \mathbf{z}_1,\dots, \mathbf{z}_{n}, \bphi),
\end{align}
where each latent layer depends on some or all of the parameters that appear upstream in the simulation. Typical examples of latent variables appearing in cosmological forward simulations might be the amplitudes and phases of the initial potential perturbations in the Universe, the (true) redshifts and other physical properties of galaxies in a survey, realizations of stochastic foregrounds or detector artifacts, etc.

In the likelihood-free set-up we aim to fit a flexible model for $p(\t | \bphi)$ to a set of simulated pairs $\{\bphi, \t\}$. Crucially, the target $p(\t | \bphi)$ is implicitly (by construction) marginalized over all of the local latent variables, since
\begin{align}
p(\t | \bphi)\equiv \int p(\mathbf{z}_1 | \bphi) & p(\mathbf{z}_2 | \mathbf{z}_1, \bphi)\dots p(\mathbf{z}_n | \mathbf{z}_1,\dots, \mathbf{z}_{n-1}, \bphi) \nonumber \\
&\times p(\t | \mathbf{z}_1,\dots, \mathbf{z}_{n}, \bphi) \,d\mathbf{z}_1d\mathbf{z}_2\dots d\mathbf{z}_n.
\end{align}
This means that if we just re-cast the nuisance parameters as an additional intermediate latent layer in the forward simulations, they will implicitly be marginalized over in the subsequent likelihood-free inference. This is easy: factorizing the prior over interesting and nuisance parameters as $p(\bphi) = p(\btheta)p(\boldsymbol{\eta} | \btheta)$,
then data-interesting parameter pairs $\{\btheta, \t\}$ can be generated from simulations as,
\begin{align}
\label{nuisance_sims}
&\btheta \leftarrow q(\btheta),\nonumber \\
&\boldsymbol{\eta} \leftarrow p(\boldsymbol{\eta} | \btheta), \nonumber \\
&\data \leftarrow \mathrm{simulator}(\data | \bphi, \boldsymbol{\eta}), \nonumber \\
&\t = \mathrm{compression}(\data).
\end{align}
These data-interesting parameter pairs $\{\btheta, \t\}$ can then be fit with a conditional density estimator for $p(\t | \btheta)$, implicitly marginalizing over the now local latent nuisance parameters $\boldsymbol{\eta}$. Note that no further approximations have been made in changing the target density from $p(\t | \btheta, \boldsymbol\eta)$ to the nuisance-marginalized $p(\t | \btheta)$. Uncertain nuisance parameters and their degeneracies with the interesting parameters are fully accounted for in the forward simulations Eq. \eqref{nuisance_sims}, so the estimated $p(\t | \btheta)$ will converge to the marginalized likelihood $\int p(\t | \btheta, \boldsymbol\eta)p(\boldsymbol\eta | \btheta) d\boldsymbol\eta$.

The nuisance-marginalized inference task has hence been reduced from inferring a $p$-dimensional density\footnote{Assuming there are $p$ compressed summaries $\t\in\mathbb{R}^p$, ie., one per parameter. This will be the case if either the score of an approximate likelihood or an information maximizing neural network is used for the summary statistics (see \S \ref{sec:compression}; \citealp{Alsing2018delfi, Charnock2018}).} as a function of $p$ parameters $p(\t | \btheta, \boldsymbol{\eta})$, to inferring a $p$-dimensional density as a function of only the $n = p - m$ interesting parameters $p(\t | \btheta)$. When the number of nuisance parameters is even modestly large, this already gives a major reduction in the computational complexity of the likelihood-free inference task.
\section{Nuisance parameter hardened data compression}
\label{sec:hardened}
Having successfully reduced the number of parameters that must be fit for in the inference step (\S \ref{sec:nuisance_latent}), the goal of this section is to reduce the number of data summaries that need to be considered, from $\t\in\mathbb{R}^p$ to some smaller set $\bar{\t}_\theta\in\mathbb{R}^n$ -- one per interesting parameter.

Taking either the (approximate) score or a trained \textsc{imnn} as the compression scheme, each compressed summary $t_{\phi_i}$ is constructed to be as informative as possible about the corresponding parameter $\phi_i$. However, degeneracies between parameters mean that each summary will also in general be sensitive to all other parameters. Our goal is to find a reduced set of ``nuisance hardened" summary statistics $\bar{\t}_\theta\in\mathbb{R}^n$ that are asymptotically optimal summaries for the $n$ interesting parameters (preserving the Fisher information) whilst having their sensitivity to the nuisance parameters projected out.

For a given likelihood $\L$, the Fisher information maximizing summaries for $n$ interesting parameters that are insensitive to some $m$ nuisances is just the score of the likelihood marginalized over the nuisance parameters, evaluated at some fiducial parameter values. We use this as our guiding principle for deriving nuisance hardened summary statistics. In the short derivation below we will work in the regime where the likelihood-function is (assumed) known: for likelihood-free applications, the derived result can be applied to the approximate methods for estimating the score described in \S \ref{sec:compression}, or \textsc{imnn} compression.

Nuisance hardened summaries can be obtained given some likelihood-function $\L$ as follows. Firstly, we Taylor expand the log-likelihood to second order in the parameters about some expansion point $\boldsymbol\phi_*$,
\begin{align}
\label{taylor}
\L(\bphi) &= \L_* + \t^\mathrm{T}\delta\bphi - \f{1}{2}\delta\bphi^\mathrm{T} \mathbf{J}\,\delta\bphi + \mathcal{O}(3)\nonumber \\
&\approx \L_* + \t^\mathrm{T}\delta\bphi - \f{1}{2}\delta\bphi^\mathrm{T} \mathbf{F}\,\delta\bphi,
\end{align}
where $\t\equiv\nabla\L_*$ is the score and $\mathbf{J}\equiv-\nabla\nabla^\mathrm{T}\L_*$ the observation matrix, which we have replaced by its expectation value -- the Fisher matrix $\mathbf{F} = -\langle\nabla\nabla^\mathrm{T}\L_*\rangle$ -- in the second line. Gradients are defined as $\nabla\equiv(\nabla_{\btheta}, \nabla_{\boldsymbol{\eta}})$.

Next, we marginalize the approximate likelihood over the nuisance parameters. In the quadratic expansion of the log-likelihood this is just a Gaussian integral, which after some algebra (dropping $\btheta$-independent terms) gives:
\begin{align}
\bar{\L}(\btheta)  &= \ln\left(\int e^{\L_* + \t^\mathrm{T}\delta\bphi - \f{1}{2}\delta\bphi^\mathrm{T} \mathbf{F}\,\delta\bphi}d\boldsymbol{\eta}\right) \nonumber \\
&= \L_* + \t_{\btheta}^\mathrm{T}\delta\btheta - \f{1}{2}\delta\btheta^\mathrm{T} \mathbf{F}_{\btheta\btheta}\,\delta\btheta \nonumber \\
&\;\;\;\;\;\;\;\;\;\;\;\;\;\;\;\;+\f{1}{2}(\mathbf{F}_{\boldsymbol{\eta}\btheta}\delta\btheta - \t_{\boldsymbol{\eta}})^\mathrm{T}\mathbf{F}^{-1}_{\boldsymbol{\eta}\boldsymbol{\eta}}(\mathbf{F}_{\boldsymbol{\eta}\btheta}\delta\btheta - \t_{\boldsymbol{\eta}})
\end{align}
where $\t_{\btheta}\equiv\nabla_{\btheta}\L_*$, $\t_{\boldsymbol{\eta}}\equiv\nabla_{\boldsymbol{\eta}}\L_*$, and the various blocks of the Fisher matrix are defined from:
\begin{align}
\mathbf{F} \equiv \left(\begin{array}{@{}cc} 
\mathbf{F}_{\btheta\btheta} & \mathbf{F}_{\btheta\boldsymbol{\eta}} \\
\mathbf{F}_{\boldsymbol{\eta}\btheta} & \mathbf{F}_{\boldsymbol{\eta}\boldsymbol{\eta}} 
\end{array}\right).
\end{align}
Finally, we obtain the nuisance-hardened summaries as the score of the approximate nuisance-marginalized likelihood $\bar{\t}_{\btheta} \equiv \nabla_{\btheta}\bar{\L}_*$, giving:
\begin{align}
\label{projection}
\bar{\t}_{\btheta} =  \t_{\btheta} - \mathbf{F}_{\btheta\boldsymbol{\eta}}\mathbf{F}^{-1}_{\boldsymbol{\eta}\boldsymbol{\eta}}\t_{\boldsymbol{\eta}}.
\end{align}
This is the main result of this section: Eq. \eqref{projection} tells us how to form nuisance-hardened summaries $\bar{\t}_{\btheta}$ from score statistics $\t$, where the only extra ingredient required for projecting out the nuisance parameter sensitivities is the Fisher matrix.

The projection in Eq. \eqref{projection} makes good intuitive sense; expected covariances $\mathbf{F}_{\btheta\boldsymbol{\eta}}$ between interesting and nuisance parameters are projected out, weighted by the expected marginal variances of the nuisance parameters $\mathbf{F}^{-1}_{\boldsymbol{\eta}\boldsymbol{\eta}}$.

How successfully the projection in Eq. \eqref{projection} removes the nuisance parameter dependence of the summaries depends on how well the approximations in Eq. \eqref{taylor} are satisfied. In the asymptotic (or linear-model) limit where the likelihood is Gaussian in the parameters, the projection and hence nuisance-hardening will be exact. When the likelihood is non-Gaussian in the parameters or there are non-linear degeneracies between interesting and nuisance parameters, the projection will be approximate. In this case, as always with likelihood-free inference, the only price to pay is some loss of optimality; the nuisance-hardened summaries will be sub-optimal, but provided the same compression scheme is applied to the data and simulations no biases are introduced.

For likelihood-free applications, Eq. \eqref{projection} can be applied to approximate score summaries (see \S \ref{sec:compression}). Alternatively, the projection in Eq. \eqref{projection} can also be straightforwardly applied to information maximizing neural networks. In maximizing the Fisher information, \textsc{imnn}s implicitly learn the score of the (true) likelihood some (non-linear) transformed data vector. The trained \textsc{imnn} gives the compressed summaries $\t$ and the Fisher information matrix $\mathbf{F}$ as output, to which Eq. \eqref{projection} can be applied directly to project out the nuisance parameters.

Note that the result in Eq. \eqref{projection} was also obtained by a different route in \citet{Zablocki2016}. In that work they sought the linear combinations of (Gaussian) data that maximize the Fisher information for each parameter in turn, whilst constrained to be insensitive (to leading order in the expectation) to all other parameters. Our results show that the \citet{Zablocki2016} result can be interpreted as the score-compression of any likelihood, marginalized over nuisance parameters in the Laplace approximation.

With the nuisance-hardened summary statistics in hand and nuisances re-cast as local latent variables (\S \ref{sec:nuisance_latent}), the nuisance marginalized inference task is now reduced to learning an $n$-dimensional density $p(\bar{\t}_{\btheta} | \btheta)$ as a function of the $n$ interesting parameters only, irrespective of the presence or number of additional nuisance parameters to be marginalized over. For even a modest number of nuisance parameters this represents a significant reduction in computational complexity compared to inferring the higher-dimensional $p(\t | \btheta, \boldsymbol\eta)$ and marginalizing over $\boldsymbol\eta$ a posteriori.
\subsection{Recovering an approximate nuisance parameter posterior}
It is possible to construct an approximate marginal nuisance parameter posterior \emph{a posteriori}, from the nuisance hardened analysis. The nuisance posterior can be approximated as follows:
\begin{align}
\label{blackwell-rao}
    p(\boldsymbol\eta | \data) &= \int p(\boldsymbol\eta | \btheta, \data)p(\btheta | \data)d\btheta, \nn \\
    &\approx \int e^{-\f{1}{2}(\boldsymbol\eta - \boldsymbol\mu_{\boldsymbol\eta}(\btheta))^\dagger \left(\mathbf{F}^{-1}_{\boldsymbol{\eta}\boldsymbol{\eta}} - \mathbf{F}^{-1}_{\boldsymbol{\eta}\btheta}\left[\mathbf{F}^{-1}_{\btheta\btheta}\right]^{-1}\mathbf{F}^{-1}_{\btheta\boldsymbol\eta}\right)^{-1}(\boldsymbol\eta - \boldsymbol\mu_{\boldsymbol\eta}(\btheta))}\nonumber \\
    &\hspace{17em}\times p(\btheta | \bar\t_{\btheta})d\btheta,
\end{align}
where in the second line we replace the marginal interesting-parameter posterior by the nuisance hardened version, and take a Laplace approximation for the nuisance conditional $p(\btheta, \boldsymbol\eta | \data)$ about $(\btheta_*,\boldsymbol\eta_*)$ (cf., Eq. \ref{taylor}). The conditional mean in the Laplace approximation term is given by
\begin{align}
\boldsymbol\mu_{\boldsymbol\eta}(\btheta) = \hat{\boldsymbol\eta} + \mathbf{F}^{-1}_{\boldsymbol{\eta}\btheta}\left[\mathbf{F}^{-1}_{\btheta\btheta}\right]^{-1}(\btheta - \hat\btheta),    
\end{align} 
where $(\hat{\boldsymbol\theta}, \hat{\boldsymbol{\eta}}) = (\btheta_*, \boldsymbol{\eta}_*) + \mathbf{F}^{-1}\mathbf{t}$ is one Newton iteration towards the maximum of the approximate likelihood used to obtain the Fisher matrix and score. The integral in Eq. \eqref{blackwell-rao} can then be approximated as a sum over samples from the interesting-parameter posterior, giving a pseudo Blackwell-Rao estimator for the nuisance parameter posterior.

While likelihood-free inference using nuisance-hardened summaries is robust in the sense that approximations made in the compression/nuisance projection steps cannot bias the parameter inferences, the same cannot be said of the above approximate nuisance parameter posterior in Eq. \eqref{blackwell-rao}. The location and scale of the approximate nuisance posterior is determined (partly) by the approximate likelihood used for performing the compression, ie., the first term in the integral Eq. \eqref{blackwell-rao}. If the approximate likelihood used for the compression is poor, the location and scale of the nuisance posterior will be biased, and in the limit where the nuisance and interesting parameters are independent, Eq. \eqref{blackwell-rao} just gives a Gaussian approximation of the approximate likelihood. Nevertheless, Eq. \eqref{blackwell-rao} gives a rough inference of the nuisances when the approximate likelihood is reasonable, and will improve as correlation between the interesting and nuisance parameters becomes more important. 
\section{validation case I: supernovae data analysis}
\label{sec:jla}
In this section we validate likelihood-free inference of nuisance marginalized posteriors on a simple case: inferring cosmological parameters from supernovae data in the presence of systematics.

Supernova data provide a great opportunity for likelihood-free methods, since the data are impacted by a large number of systematic biases and selection effects that need to be carefully accounted for to obtain robust cosmological parameter inferences.

Here, for the purpose of validation, we perform a simple analysis of the JLA data \citep{Betoule2014} under assumptions that allow us to compare the likelihood-free results against an exact (known) likelihood. The set-up is identical to \citet{Alsing2018delfi}, which we review briefly below.
\subsection{JLA data and model}
The JLA sample is comprised of $740$ type Ia supernovae for which we have estimates for the apparent magnitudes $m_\mathrm{B}$, redshifts $z$, color at maximum-brightness $C$ and stretch $X_1$ parameters characterizing the lightcurves. We take the data vector to be the vector of estimated apparent magnitudes $\data = (\hat{m}_\mathrm{B}^1, \hat{m}_\mathrm{B}^2,\dots,\hat{m}_\mathrm{B}^M)$, with uncertainties in the redshift, color and stretch approximately accounted for in the covariance matrix of the observed apparent magnitudes \citep{Betoule2014}.

We assume that the apparent magnitudes of type Ia supernovae depend on the luminosity distance to the source at the given redshift $D^*_\mathrm{L}(z)$, a reference absolute magnitude for type Ia supernovae, and calibration corrections for the stretch $X_1$ and color at maximum-brightness $C$ \citep{Tripp1998},
\begin{align}
\label{apparent_mag}
m_\mathrm{B} = 5\mathrm{log}_{10}\left[\f{D^*_\mathrm{L}(z ; \btheta)}{10\mathrm{pc}}\right] &- \alpha X_1 + \beta C \nonumber \\
& + M_\mathrm{B} + \delta M\,\Theta(M_\mathrm{stellar} - 10^{10}M_\odot),
\end{align}
where $\btheta$ are the cosmological parameters (see below), $\alpha$ and $\beta$ are calibration parameters for the stretch and color, and $\tilde{M}_\mathrm{B}$ and $\delta M$ characterize the host stellar-mass ($M_\mathrm{stellar}$) dependent reference absolute magnitude. $\Theta$ is the Heaviside function.

The cosmological model enters via the luminosity distance-redshift relation: we will assume a flat $w$CDM universe with cold dark matter (with total matter density parameter $\Omega_\mathrm{m}$) and dark energy characterized by equation-of-state $p/\rho=w_0$. The luminosity distance-redshift relation is given by
\begin{align}
D^*_\mathrm{L}(z; \btheta) = \f{(1+z)c}{100}\int_0^z \f{dz'}{\sqrt{\Omega_\mathrm{m}(1+z')^3 + (1-\Omega_\mathrm{m})(1+z')^{3(w_0+1)}}},
\end{align}
where $c$ is the speed of light in vacuum. The cosmological parameters of interest are $\btheta = (\Omega_\mathrm{m}, w_0)$, and we treat the remaining parameters $\boldsymbol\eta = (\alpha, \beta, M_\mathrm{B}, \delta M)$ as nuisances.
\subsection{Likelihood and data compression}
For this validation case we assume the data are Gaussian, 
\begin{align}
\label{jla_sampling}
\ln\,p(\data | \bphi) = -\f{1}{2}(\data - \boldsymbol\mu(\bphi))^T\cov^{-1}(\data - \boldsymbol\mu(\bphi)) - \f{1}{2}\ln |\cov|,
\end{align}
with mean given by Eq. \eqref{apparent_mag}, and we assume a fixed covariance matrix from \citet{Betoule2014} (see \citealp{Alsing2018delfi} for details of the covariance matrix).

We take compressed data summaries to be the score of the Gaussian likelihood, ie.,
\begin{align}
\t \equiv\nabla_{\bphi}\mathcal{L}_*= \nabla_{\bphi}^T\boldsymbol\mu_*\cov^{-1}(\data - \boldsymbol\mu_*),
\end{align}
where we take fiducial parameters for the expansion point $\btheta_* = (0.202, -0.748, -19.04, 0.126, 2.644, -0.0525)$\footnote{Found in a few iterations of $\hat\btheta_{k+1} = \hat\btheta_{k} + \mathbf{F}^{-1}_k\t_k$}, and `$*$' indicated evaluation at the fiducial parameters.

Projection of the nuisance parameters is performed following Eq. \eqref{projection}, giving nuisance hardened summary statistics for the interesting (cosmological) parameters:
\begin{align}
\bar{\t}_{\btheta} =  \t_{\btheta} - \mathbf{F}_{\btheta\boldsymbol{\eta}}\mathbf{F}^{-1}_{\boldsymbol{\eta}\boldsymbol{\eta}}\t_{\boldsymbol{\eta}},
\end{align}
where the Fisher information matrix is given by $\mathbf{F} = \nabla_{\bphi}\boldsymbol\mu^T \cov^{-1}\nabla_{\bphi}^T\boldsymbol\mu$.

For a full sophisticated implementation of likelihood-free inference to supernova data analysis, in practise, may use an approximate likelihood or information maximizing neural network for performing data compression.
\subsection{Simulations}
For this validation case, simulations are just draws from the (exact) assumed sampling distribution of the data, ie., drawing Gaussian data from Eq. \eqref{jla_sampling} (given parameters).
\subsection{Priors}
We assume broad Gaussian priors on the cosmological parameters $\btheta= (\Omega_\mathrm{m}, w_0)$  with mean and covariance (following \citealp{Alsing2018delfi}):
\begin{align}
\boldsymbol{\mu}_\mathrm{P,\btheta} = (0.3,\;  -0.75),\;\cov_\mathrm{P,\btheta} =
  \left( {\begin{array}{cc}
   0.4^2 & -0.24  \\
   -0.24 & 0.75^2  \\
\end{array} } \right),
\end{align}
with additional hard prior boundaries $\Omega_\mathrm{m}\in[0, 0.6]$ and $w_0\in[-1.5, 0]$.

We take independent Gaussian priors on the nuisance parameters $\boldsymbol\eta = (\alpha, \beta, M_\mathrm{B}, \delta M)$, with means $\boldsymbol{\mu}_\mathrm{P,\boldsymbol\eta} = (-19.05 ,\;   0.125,\; 2.6  ,  \;-0.05)$ and standard deviations $\boldsymbol\sigma_\mathrm{P,\boldsymbol\eta} = (0.1,\;0.25, \;0.025,\;0.05)$.
\subsection{DELFI set-up}
The density-estimation likelihood-free inference (DELFI) algorithm learns the sampling distribution of the data summaries as a function of the parameters: we parameterize the data sampling distribution by a mixture density network with three Gaussian components, and simulations are ran in sequential batches with an adaptive proposal density (following \citealp{Papamakarios2018,Alsing2019neural}). Further technical details of the network architectures and DELFI algorithm set-up used here can be found in Appendix \ref{app:delfi_setup}. We used the public DELFI implementation \textsc{pydelfi}\footnote{\url{https://github.com/justinalsing/pydelfi}} \citep{Alsing2019neural}.
\subsection{Results}
\begin{figure}
  \centering
  \includegraphics[width=8cm]{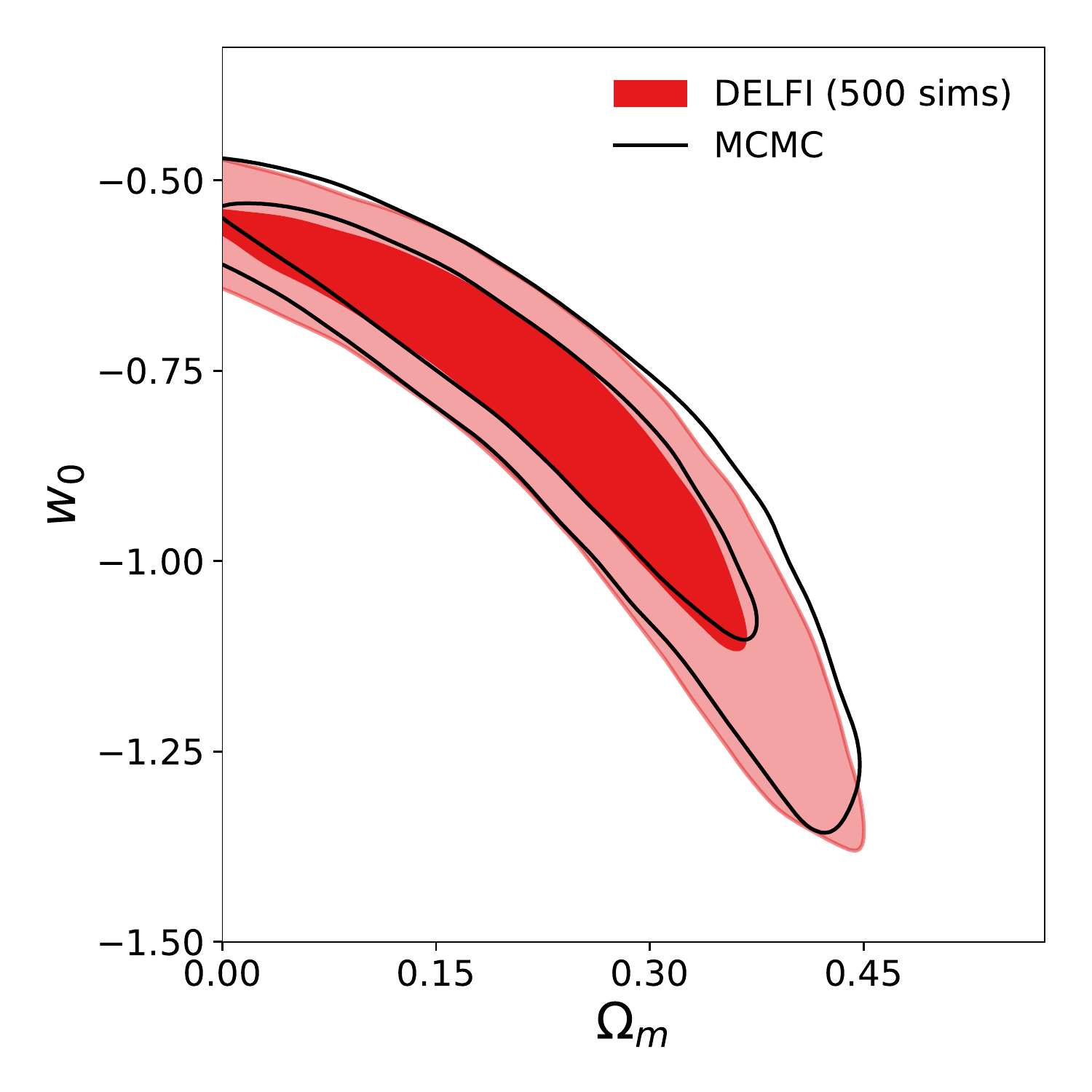}
\caption{68 and 95\% confidence contours for the nuisance marginalized posterior using direct likelihood-free inference of the marginalized posterior (trained on $500$ simulations; red), compared to MCMC sampling of the exact posterior (black). In this case, the likelihood-free approach is in excellent agreement with the exact marginal posterior.}
\label{fig:jla_comparison}
\end{figure}
\begin{figure*}
  \centering
  \subfigure[2-parameter nuisance marginalized problem]{\includegraphics[scale=0.48]{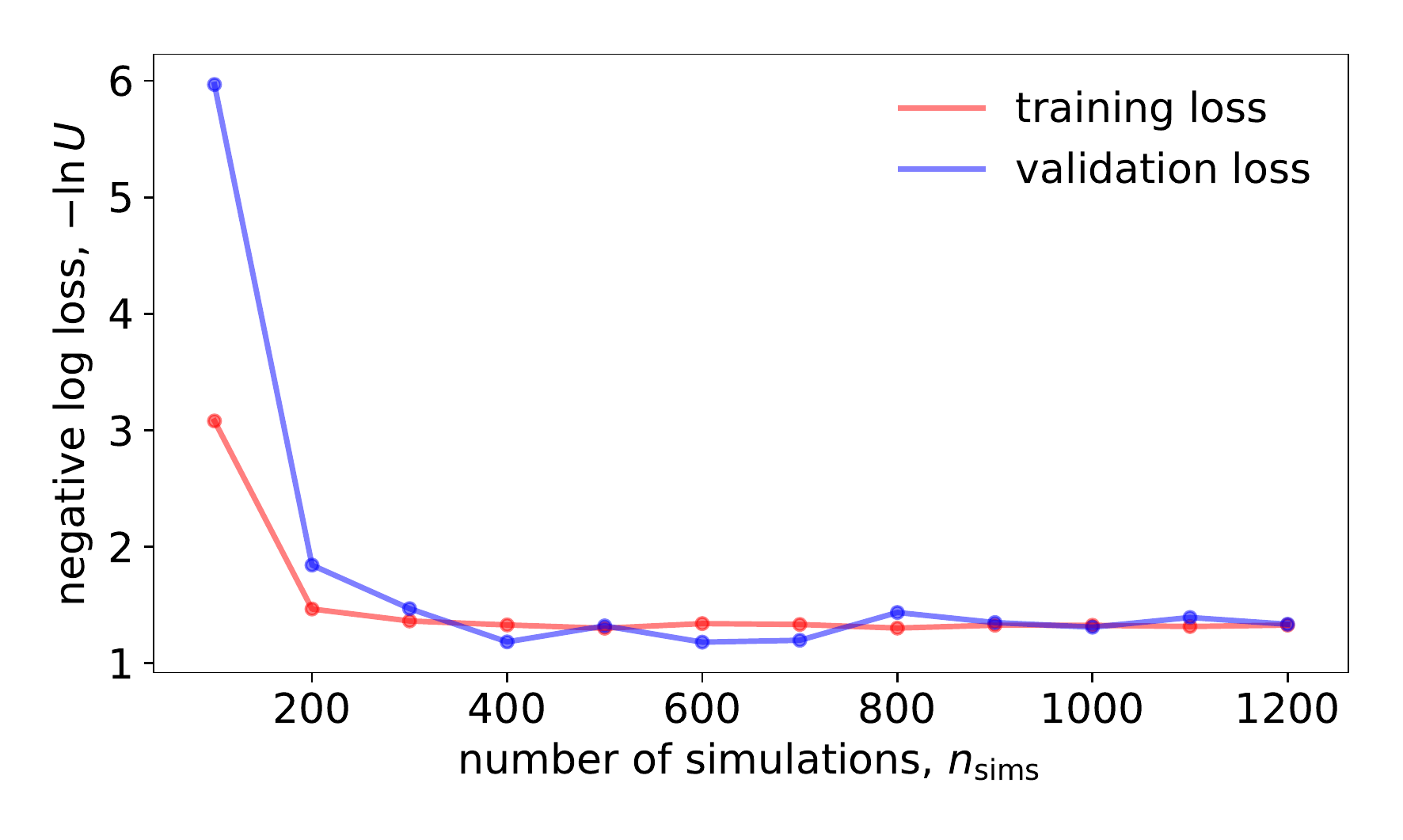}}\quad
  \subfigure[6-parameter problem (including nuisance parameters)]{\includegraphics[scale=0.48]{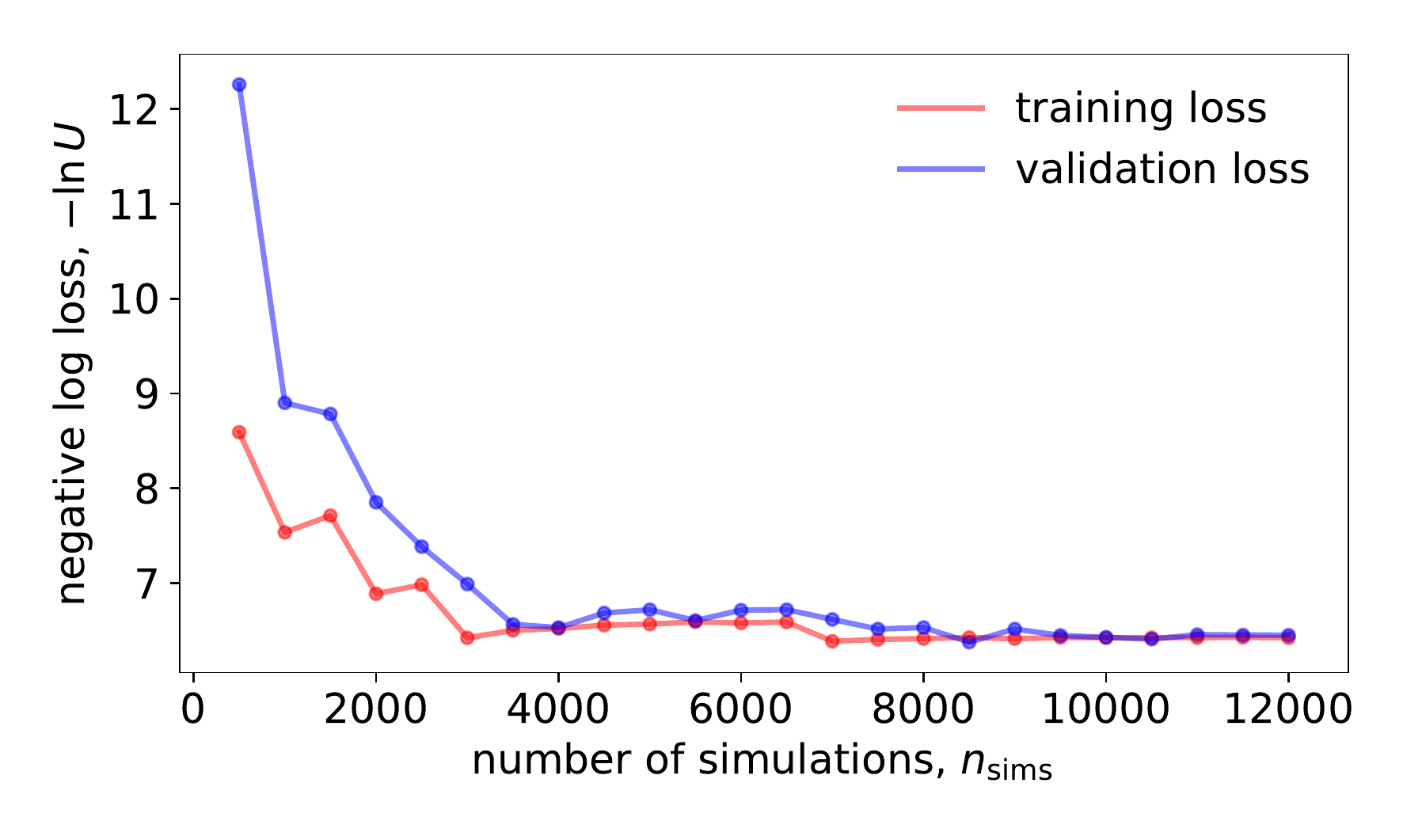}}
\caption{Convergence of the learned DELFI posterior as a function of the number of forward simulations for the 2-parameter nuisance marginalized posterior (left), and the 6-d joint nuisance-interesting parameter posterior (right). Directly inferring the nuisance marginalized posterior required roughly an order of magnitude fewer simulations compared to inferring the joint posterior over nuisances and interesting parameters first and marginalizing a posteriori.}
\label{fig:jla_convergence}
\end{figure*}
\begin{figure*}
  \centering
  \includegraphics[width=17cm]{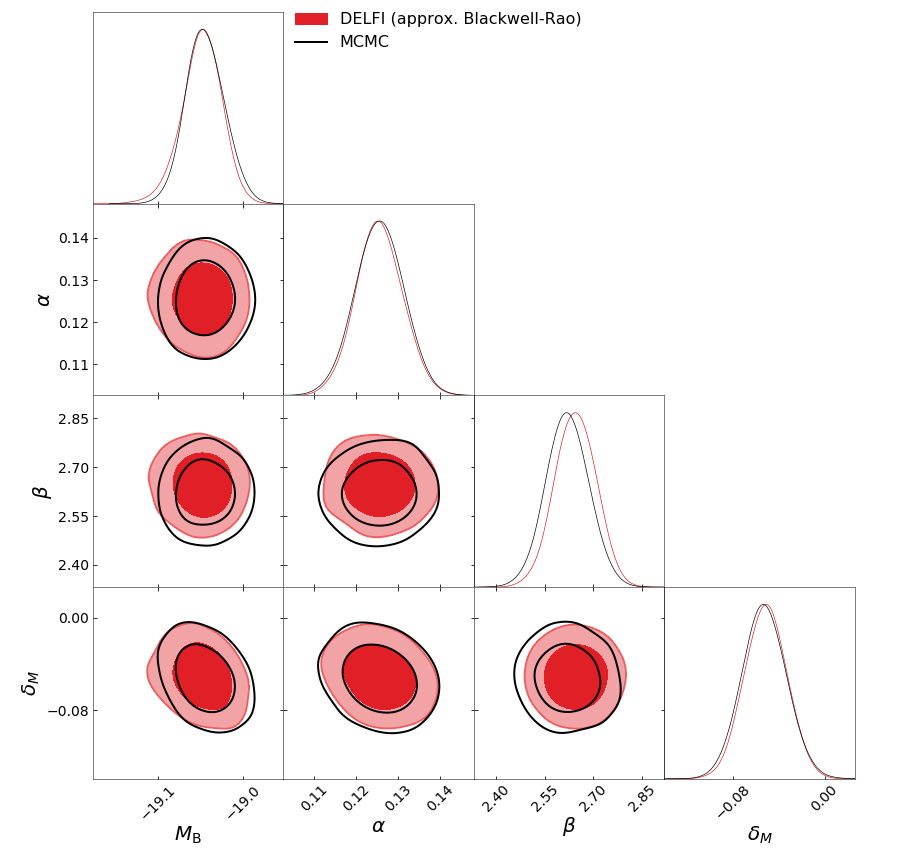}
\caption{68 and 95\% confidence contours for the approximate Blackwell-Rao estimator for the nuisance parameter posterior (red), compared to MCMC sampling of the exact posterior (black). In this case (since the assumptions underpinning the approximate Blackwell-Rao posterior are satisfied), the likelihood-free approach is in excellent agreement with the exact marginal posterior.}
\label{fig:nuisance_posterior}
\end{figure*}
We ran the JLA inference problem in two scenarios. First, we inferred the posterior for the six interesting plus nuisance parameters jointly, ie., learning the $6$-dimensional density $p(\t_{\btheta}, \t_{\boldsymbol\eta} | \btheta, \boldsymbol\eta)$ as a function of the six parameters. Secondly, we inferred the posterior for the two interesting parameters directly marginalized over the nuisances, ie., learning the 2-dimensional density $p(\bar{\t}_{\btheta} | \btheta)$ as a function of the two cosmological parameters.

In Figure \ref{fig:jla_comparison} we see that the nuisance marginalized posterior obtained in the likelihood-free setting is in excellent agreement with exact marginal posterior (obtained by sampling the full 6-parameter likelihood with MCMC). This means that for this particular example, the nuisance hardened data compression described in \S \ref{sec:hardened} is effectively lossless. This is expected if the Laplace approximation is a good approximation for integrating out the nuisance parameters, the likelihood used for the score compression is close to the true likelihood (in this validation case it is exact), and the expansion is performed close to the peak. The further one deviates from those conditions, the more lossy we can expect the compression and projection to be.

Figure \ref{fig:jla_convergence} shows the convergence of the negative log-likelihood of the simulated parameter-data summary samples under the learned neural density estimator. For this problem, inferring the nuisance marginalized posterior directly gives roughly an order of magnitude reduction in the number of simulations required: the two-parameter problem converges after a few hundred simulations, while the full six-parameter problem takes a few thousand. We note also that for the two-parameter problem, the number of simulations required here is an order of magnitude fewer than was reported for the same problem using Bayesian Optimization Likelihood-free Inference (BOLFI) in \citet{Leclercq2018} (in that work, the nuisance parameters were kept fixed rather than marginalized over as they are here). Note also that this small number of simulations is also in spite of assuming a flexible model for the sampling distribution of the data (summaries); a three component Gaussian mixture, where in this case the data are actually Gaussian.

Figure \ref{fig:nuisance_posterior} shows the recovered Blackwell-Rao estimator for the approximate nuisance parameter posterior. In this specific case, since the likelihood used to perform the compression was exact, the assumptions underpinning the Blackwell-Rao estimated nuisance posterior are satisfied and the recovered nuisance posterior approximation is hence very reasonable.
\section{validation case II: cosmic shear with photo-z systematics}
\label{sec:cosmic_shear}
In this section we demonstrate the utility of inferring nuisance marginalized posteriors in the likelihood-free setting on a second validation example: inferring cosmological parameters from cosmic shear observations, with nuisance parameters describing systematic uncertainties in the photometric redshift distributions.

Cosmic shear data are a good candidate for likelihood-free analysis, since they are impacted by a number of effects that can be included (to varying degrees) in forward simulations but are hard to build into a high-fidelity likelihood function. These include non-linear physics and baryonic feedback on small scales \citep{Rudd2008,Harnois2015}, intrinsic alignments \citep{Joachimi2015galaxy}, shape and photo-$z$ measurement systematics \citep{Massey2012,Mandelbaum2018,Salvato2018}, blending \citep{Mandelbaum2018}, reduced shear corrections \citep{Krause2010}, redshift-dependent galaxy-property distributions \citep{Kannawadi2018}, non-trivial sampling distributions for common summary statistics \citep{Sellentin2018}, etc., all of which can result in biased inferences if not carefully accounted for. Likelihood-free methods may also allow us to extract extra information from non-standard lensing observables (eg., magnification \citealp{Hildebrandt2009,VanWaerbeke2010,Hildebrandt2013,Duncan:2013bb,Heavens2013, Alsing2015}), and summary statistics such as convergence peaks \citep{Kratochvil2010,Fluri2018peaks}, bispectra \citep{Cooray2001} etc.

Here, for the purpose of validation, we construct a simplified toy model for cosmic shear data where we can validate the likelihood-free results against an exact (known) likelihood.
\subsection{Tomographic cosmic shear data and model}
The tomographic angular shear power spectra contain much of the cosmological information contained in the weak lensing distortion fields and are straightforwardly predicted from theory. For a given set of cosmological parameters $\btheta$, the predicted angular power spectra between tomographic bins $\alpha$ and $\beta$ are given by\footnote{In the Limber approximation, \citep{Limber1954}.} \citep{Kaiser1992, Kaiser1998, Hu1999, Hu2002a, Takada2004, Kitching2017},
\begin{align}
\label{limber_power}
C_{\ell,\alpha\beta}^{\gamma\gamma} &= \int \f{d\chi}{\chi^2}\;w_\alpha(\chi)w_\beta(\chi)\left[1+z(\chi)\right]^2P_\delta\left(\f{\ell}{\chi}; z(\chi)\right),
\end{align}
where cosmological parameters $\btheta$ enter via the comoving distance-redshift relation $\chi(z)$, the 3D power spectrum of matter fluctuations $P_\delta(k; \chi)$, and the lensing weight functions which are given by
\begin{align}
\label{weight}
w_\alpha(\chi)=\f{3\Omega_\mathrm{m}H_0^2}{2}\chi\int_{\chi}^{\chi_\mathrm{H}} d\chi'\;n_\alpha(\chi')\f{\chi'-\chi}{\chi'},
\end{align}
where $n_\alpha(\chi)d\chi = p_\alpha(z)dz$ is the redshift distribution for galaxies in redshift bin $\alpha$.

The tomographic redshift distributions $p_\alpha(z)$ must be estimated from photometric data and are subject to a range of systematic biases that typically must be parameterized and marginalized over (see eg., \citealp{Salvato2018} for a review). For this demonstration we parameterize systematic uncertainties in the tomographic redshift distributions as an unknown shift of the estimated distributions, ie., setting
\begin{align}
\label{pz}
p_\alpha(z) = \hat p_\alpha(z - b_\alpha),
\end{align}
where $\hat p_\alpha(z)$ is the estimated redshift distribution for tomographic bin $\alpha$ and the systematic shift parameters $b_\alpha$ -- one per tomographic bin -- constitute the nuisance parameters $\boldsymbol{\eta}$ that we ultimately want to marginalize over.

For this demonstration we will assume the data vector to be a set of estimated (noise biased) band powers,
\begin{align}
\data = (\hat{\cov}_{\mathcal{B}_1}, \hat{\cov}_{\mathcal{B}_2}, \dots, \hat{\cov}_{\mathcal{B}_K}).
\end{align}

We consider a flat-$w$CDM cosmology parameterized by $\btheta = (\sigma_8, \Omega_\mathrm{m}, \Omega_\mathrm{b}h^2, h, n_s, w_0)$.
\subsection{Likelihood and data compression}
For this validation case study, we assume that the data (band powers) are independently Wishart distributed,
\begin{align}
\label{cs_like}
p(\data | \btheta, \boldsymbol\eta) = \prod_{\mathrm{bands,\,k}} \mathcal{W}\left(\hat\cov_{\mathcal{B}_k}\;|\; \cov_{\mathcal{B}_k}(\btheta, \boldsymbol\eta), \nu_{\mathcal{B}_k}\right),
\end{align}
where the elements of the expected band powers are given by
\begin{align}
C_{\mathcal{B}_k, ij} = \sum_{\ell\in\mathcal{B}_k} (C_{\ell, ij}^{\gamma\gamma}(\btheta, \boldsymbol\eta) + N_{\ell, ij})/\nu_{\mathcal{B}_k}.
\end{align}
$N_{\ell,ij}$ is the shape noise power spectrum (from intrinsic random galaxy ellipticities) and $\nu_{\mathcal{B}_k}$ is the total number of modes contributing to band $k$. We will assume isotropic shape noise $N_{\ell,ij} = \sigma_e^2/\bar{n}_i\delta_{ij}$ with intrinsic ellipticity variance $\sigma_e^2$ and mean galaxy number density $\bar{n}_i$ in tomographic bin $i$. We will assume $\nu_{\mathcal{B}_k} = f_\mathrm{sky}\sum_{\ell\in\mathcal{B}_k}(2\ell+1)$ modes per band to approximately mimick partial sky-fraction coverage $f_\mathrm{sky}$.

We take our compressed summaries as the score of the Wishart likelihood in Eq. \eqref{cs_like},
\begin{align}
\t\equiv\nabla_{\bphi}\mathcal{L}_* = \sum_k \f{\nu_{\mathcal{B}_k}}{2}\mathrm{tr}\big[\cov_{\mathcal{B}_k*}^{-1}&\nabla\cov_{\mathcal{B}_k*}\cov_{\mathcal{B}_k*}^{-1}\hat\cov_{\mathcal{B}_k}) - \cov_{\mathcal{B}_k*}^{-1}\nabla\cov_{\mathcal{B}_k*}\big],
\end{align}
where `$*$' indicates evaluation about fiducial parameters $\btheta_*$ which we take to be $\btheta_* = (0.8, 0.3, 0.05, 0.7, 0.96, -1)$.

Projection of the nuisance parameters is performed following Eq. \eqref{projection} as usual, giving nuisance hardened summary statistics:
\begin{align}
\bar{\t}_{\btheta} =  \t_{\btheta} - \mathbf{F}_{\btheta\boldsymbol{\eta}}\mathbf{F}^{-1}_{\boldsymbol{\eta}\boldsymbol{\eta}}\t_{\boldsymbol{\eta}},
\end{align}
where the Fisher matrix in this case is given by,
\begin{align}
\mathrm{F}_* = -\langle\nabla_{\btheta}\nabla_{\btheta}^T\mathcal{L}\rangle_* = \sum_k \f{\nu_{\mathcal{B}_k}}{2}\mathrm{tr}\big[\cov_{\mathcal{B}_k*}^{-1}&\nabla_{\btheta}\cov_{\mathcal{B}_k*}\cov_{\mathcal{B}_k*}^{-1}\nabla_{\btheta}^T\cov_{\mathcal{B}_k*}\big].
\end{align}

For a full sophisticated implementation of likelihood-free inference to cosmic shear on real data, an approximate likelihood or information maximizing neural network may be used for performing data compression.
\subsection{Simulations}
\label{sec:mock_data}
In this simple validation case we simulate data (given parameters) by drawing band powers as Wishart random variates, following Eq. \eqref{cs_like}.

We assume a survey set-up similar to the upcoming ESA \emph{Euclid} survey: $15,000$ square degrees with a mean galaxy number density of $\bar{n}=30\,\mathrm{arcmin}^{-2}$, an overall galaxy redshift distribution $n(z) \propto z^2\mathrm{exp}\left[-(1.41z/z_m)^{1.5}\right]$ with a median redshift $z_m = 0.9$, Gaussian photo-$z$ uncertainties with standard deviation $\sigma_z = 0.05*(1+z)$, and ten tomographic bins with equal mean galaxy number density per bin. Modes are binned into ten logarithmically spaced bands between $\ell=10$ and $\ell=3000$.

The data for this demonstration are simulated following Eq. \eqref{cs_like} assuming values for the cosmological parameters (based on \citealp{Planck2018}) $\sigma_8 = 0.811$, $\Omega_\mathrm{m} = 0.315$, $\Omega_\mathrm{b}h^2 = 0.0224$, $h = 0.674$, $n_s = 0.965$, $w_0=-1.03$, and systematics parameters $b_i = 0\;\forall i$.
\subsection{Priors}
\label{sec:priors}
For the interesting parameters $\btheta = (\sigma_8, \Omega_\mathrm{m}, \Omega_\mathrm{b}h^2, h, n_s, w_0)$ we assume Gaussian priors with means $\boldsymbol{\mu}_{\btheta} = (0.8, 0.3, 0.0224, 0.674, 0.96, -1.0)$, standard deviations $\boldsymbol{\sigma}_{\btheta} = (0.1, 0.1, 0.00015, 0.005, 0.3, 0.3)$, and hard parameter limits $\sigma_8 \in \left[0.4, 1.2\right]$,  $\Omega_\mathrm{m} \in \left[0, 1\right]$, $\Omega_\mathrm{b}h^2 \in \left[0, 0.1\right]$, $h \in \left[0.4, 1\right]$, $n_\mathrm{s} \in \left[0.7, 1.3\right]$. The tight priors on $\Omega_\mathrm{b}h^2$ and $h$ (which are poorly constrained by weak lensing data) are taken from \citep{Planck2018}. 

For the nuisance parameters $\{b_i\}$ we assume independent zero-mean Gaussian priors with standard deviation $0.05$ and hard limits $b_i \in \left[-0.1, 0.1\right]$.
\subsection{DELFI set-up}
\label{sec:delfi_setup}
As in \S \ref{sec:jla}, the DELFI algorithm learns the sampling distribution of the data summaries as a function of the parameters. As before we parameterize the sampling distribution by a mixture density network with three Gaussian components, and simulations are ran in sequential batches with an adaptive proposal density (using the public code \textsc{pydelfi} \citealp{Alsing2019neural}; see also Appendix \ref{app:delfi_setup}).
\begin{figure}
\centering
\includegraphics[width = 8cm]{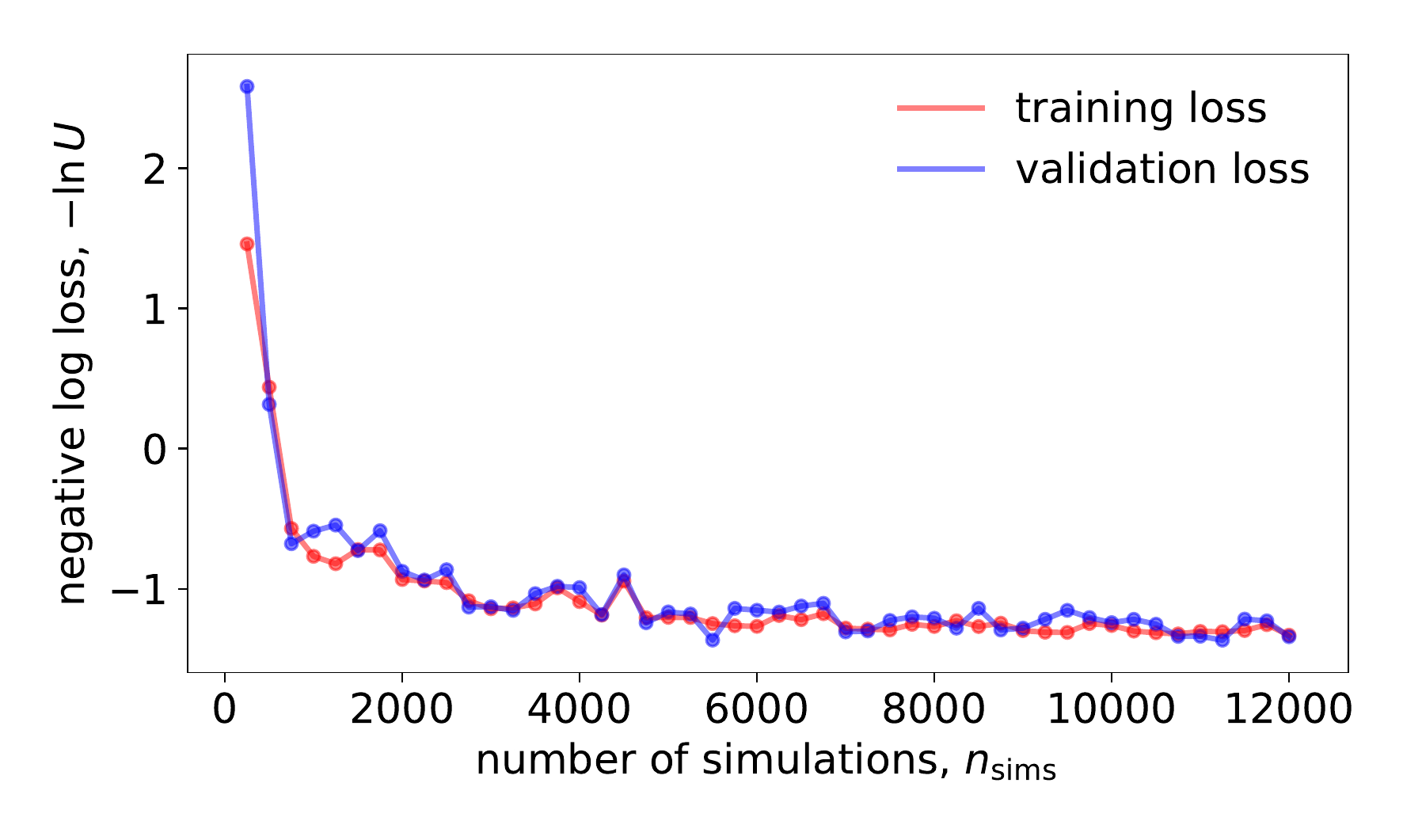}
\caption{Convergence of the learned DELFI posterior as a function of the number of forward simulations for the 6-parameter cosmic shear posterior, implicitly marginalized over ten additional nuisance parameters characterizing photo-$z$ systematics. Convergence is achieved after a few thousand simulations.}
\label{fig:convergence}
\end{figure}
\begin{figure*}
\centering
\includegraphics[width = 17.5cm]{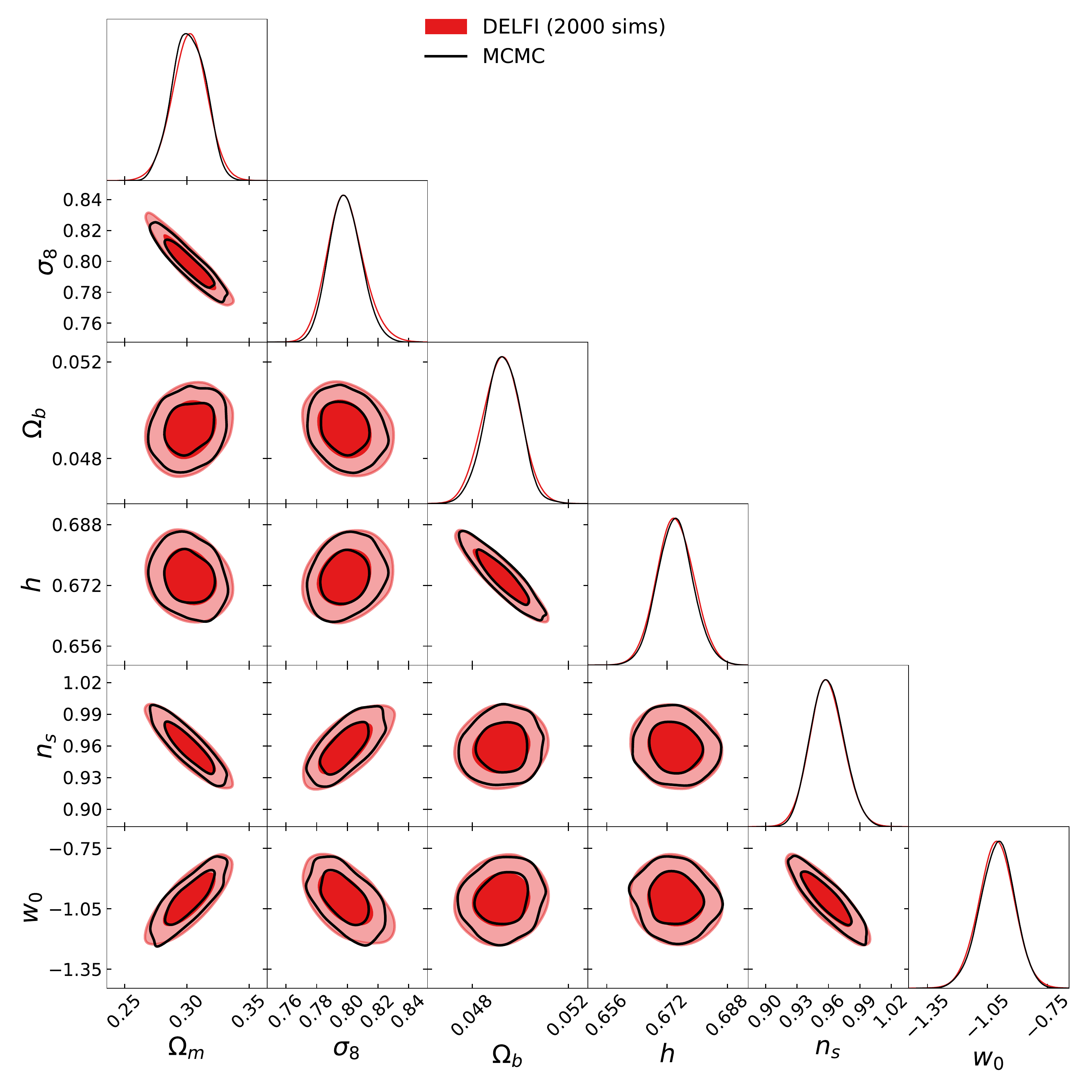}
\caption{Red: Inferred posterior marginalized over ten photo-$z$ nuisance parameters obtained with DELFI, from $2000$ forward simulations. Contours represent 68 and 95\% credible regions of the 2D marginals. Black: contours and marginals for the same photo-$z$ marginalized posterior, sampled using MCMC. The nuisance parameter hardened data compression and likelihood-free inference have successfully preserved the information content of the data, yielding a high-fidelity nuisance marginalized posterior.}
\label{fig:contours}
\end{figure*}
\subsection{Results}
%

We ran the cosmic shear inference task described above, inferring the six interesting cosmological parameters, implicitly marginalizing over the ten additional nuisances describing the photo-$z$ systematics.

Figure \ref{fig:convergence} shows the convergence of the negative log-likelihood of the simulated parameter-data summary samples under the learned neural density estimator, as a function of the number of forward simulations. We see that convergence is achieved after $\mathcal{O}(10^3)$ forward simulations, irrespective of the presence of ten additional nuisance parameters that are implicitly marginalized over here.

Figure \ref{fig:contours} shows the recovered nuisance marginalized posterior for the six cosmological parameters using likelihood-free inference (red) versus MCMC sampling of the exact nuisance plus interesting parameter problem and marginalizing a posteriori. We see that again in this case, the nuisance hardened data compression has successfully retained essentially all of the information content of the data with respect to the interesting parameters, yielding a high-fidelity nuisance marginalized (likelihood-free) posterior inference.
\section{Conclusions}
\label{sec:conclusions}
We have shown that nuisance marginalized posteriors can be inferred directly in the likelihood-free paradigm, massively reducing the number of simulations required to perform likelihood-free inference in the presence of nuisance parameters. This opens up likelihood-free methods to problems with expensive forward simulations, a small number of parameters of interest, but a large number of additional nuisance parameters; a common scenario in cosmological data analysis.

To achieve fast direct inference of nuisance marginalized posteriors, we do the following: first, we showed that nuisance parameters can easily be re-cast as local latent variables and hence implicitly marginalized over in the likelihood-free inference framework (\S \ref{sec:nuisance_latent}). Secondly, we derived a scheme for compressing $N$ data down to $n$ informative summaries -- one per parameter of interest -- such that the compressed summaries (asymptotically) retain the Fisher information content of the data for the interesting parameters whilst being ``hardened" to the nuisance parameters. This means that just $n$ data summaries can be used for likelihood-free inference, regardless of the number of nuisance parameters.

The result is that the complexity of the inference task -- and hence the number of simulations required for performing inference -- is dictated only by the number of interesting parameters considered in the problem.
\section*{Acknowledgements}
This work is supported by the Simons Foundation. Justin Alsing was partially supported by the research project grant ``Fundamental Physics from Cosmological Surveys" funded by the Swedish Research Council (VR) under Dnr 2017-04212. Benjamin Wandelt acknowledges support by the Labex Institut Lagrange de Paris (ILP) (reference ANR-10-LABX-63) part of the Idex SUPER, and received financial state aid managed by
the Agence Nationale de la Recherche, as part of the programme Investissements d'avenir
under the reference ANR-11-IDEX-0004-02.
	



\bibliographystyle{mnras}
\bibliography{nuisance}



\appendix

\section{DELFI algorithm set-up}
\label{app:delfi_setup}
We parameterize the sampling distribution of the data summaries $p(\t | \btheta)$ as a mixture density network. Simulations are run in batches and the mixture network is re-trained after each new batch of simulations. In the first cycle, parameters for performing forward simulations are drawn from a broad Gaussian centered on the prior mean and with covariance equal to nine times the (estimated) Fisher matrix. For subsequent batches of simulations, parameters are drawn from the geometric mean of the current posterior estimate from the mixture network and the prior\footnote{This proposal density is inspired by the optimal parameter proposal for Approximate Bayesian Computation \citep{Alsing2018optimal}.}.

The neural networks are trained using the stochastic gradient optimizer \textsc{adam} \citep{adam}, with a batch-size of one tenth of the training set during each re-training cycle and a learning rate of $0.001$. Over-fitting is prevented during training using early-stopping; during each re-training cycle, 10\% of the training set is set aside for validation and training is terminated when the validation-loss does not improve after $20$ epochs.

We set the size of the mixture network architecture and number of new simulations per training cycle depending on the number of parameters to be inferred. We use the following rule of thumb for scaling the architecture and simulation batches: all of the networks have two hidden layers with $5\times$ as many units as parameters to be inferred, and simulations are run in batches of $50\times$ the number of parameters in the problem. Of course, this is a very rough heuristic and careful application of neural density estimators should involve a cross-validation search over network architectures.
%

\bsp	
\label{lastpage}
\end{document}